\documentclass[conference]{IEEEtran}
\IEEEoverridecommandlockouts

\usepackage{epsfig,amsmath,amssymb,epsf,cite,algorithm,algorithmic}
\usepackage{graphicx}

\newtheorem{theorem}{Theorem}

\newtheorem{lemma}{Lemma}

\newtheorem{definition}{Definition}

\usepackage{stmaryrd}
\usepackage{amssymb}
\usepackage{tipa}
\usepackage{setspace}

\def\b0{{\pmb{0}}}

   \def\bx{{\boldsymbol{x}}}
\def\be{{\boldsymbol{e}}}  \def\bs{{\boldsymbol{s}}} 
 \def\bm{{\boldsymbol{m}}}

   \def\bX{{\boldsymbol{X}}}
  \def\bS{{\boldsymbol{S}}} \def\bY{{\boldsymbol{Y}}}
   \def\bZ{{\boldsymbol{Z}}}

   \def\bxx{{\underline{\boldsymbol{x}}}}
  \def\bss{{\underline{\boldsymbol{s}}}}

   \def\bXX{{\underline{\boldsymbol{X}}}}
  \def\bSS{{\underline{\boldsymbol{S}}}}

\begin{document}
%
\title{Many-Access Channels: The Gaussian Case with Random User Activities }

\author{
  \authorblockN{Xu Chen and Dongning Guo \\}
\authorblockA{Department of Electrical Engineering \& Computer Science \\
Northwestern University, Evanston, IL, 60208, USA
\vspace{-1ex}
}
\thanks{This material is based upon work supported in part by the National Science Foundation under Grant Nos.~1231828 and~1018578.}

}

%


\maketitle

\begin{abstract}
Classical multiuser information theory studies the fundamental limits of models with a fixed (often small) number of users as the coding blocklength goes to infinity. This work proposes a new paradigm, referred to as {\em many-user information theory}, where the number of users is allowed to grow with the blocklength. This paradigm is motivated by emerging systems with a massive number of users in an area, such as machine-to-machine communication systems and sensor networks. The focus of the current paper is the many-access channel model, which consists of a single receiver and many transmitters, whose number increases unboundedly with the blocklength. Moreover, an unknown subset of transmitters may transmit in a given block and need to be identified. A new notion of capacity is introduced and characterized for the Gaussian many-access channel with random user activities. The capacity can be achieved by first detecting the set of active users and then decoding their messages.
\end{abstract}

\section{Introduction}
\label{sec:intro}

The prevailing models in multiuser information theory assume a fixed (usually small) number of users, where fundamental limits as the coding blocklength goes to infinity are studied. Even in the large-system analysis of multiuser systems~\cite{verdu1999spectral,guo2005randomly}, the blocklength is sent to infinity before the number of users is sent to infinity.\footnote{The same can be said of the many-user broadcast coding strategy for the point-to-point channel proposed in~\cite{shamai1997broadcast}.} In some sensor networks and emerging machine-to-machine communication systems, a massive and ever-increasing number of wireless devices may need to share the spectrum in a given area. This motivates us to rethink the assumptions of fixed number of users. Here we propose a new {\em many-user paradigm}, where the number of users is allowed to increase without bound with the blocklength. The only existing model of this nature is found in \cite{chang1979coding}, in which the authors sought for uniquely-decodable codes for a noiseless binary adder channel with the number of users increasing with the blocklength.

In particular, we introduce the many-access channel~(MnAC) to model systems consisting of a single receiver and many transmitters, the number of which is comparable to or even larger than the blocklength. We study the asymptotic regime where the number of transmitting devices ($k_n $) increases with the blocklength ($n$). The model also accommodates the random access nature of the devices, allowing each transmitter to be active with certain probability in each block.

The capacity of conventional multiaccess channel model has been extensively studied \cite{ahlswede71,liao72,gallager85}. The most familiar capacity-achieving scheme relies on the fact that joint typicality holds with high probability with the number of transmitters fixed and the blocklength growing to infinity. This argument, however, does not directly apply to models where the number of users scales with the blocklength. Specifically, joint typicality requires the simultaneous convergence of the empirical joint entropy of every subset of the input and output random variables to the true joint entropy. Even though convergence holds for every subset due to the law of large numbers, the asymptotic equipartition property is not guaranteed because the number of those subsets increases exponentially with the number of variables.

In this paper, the capacity of Gaussian many-access channels is established.  In particular, the capacity of MnAC with random user activities is shown to be in the form of the capacity of MnAC with known user activities less some penalty. An achievable scheme is proposed, where each transmission consists of a signature that identifies the user, followed by a message-bearing codeword. A two-stage decoding scheme is shown to achieve the capacity. The first stage identifies the set of active users based on the superposition of their unique signatures. (This is in fact a compressed sensing problem \cite{zhang2013neighbor}.) The second stage decodes the messages from the identified active users. This scheme provides an intuitive interpretation of the capacity of MnAC: The difference between the MnAC sum capacity with and without active user information is essentially the entropy of user activities.

In previous work \cite{chen2013gaussian}, we studied the capacity of the Gaussian MnAC where all users are always active and the number of users is sublinear in the blocklength, i.e., $k_n=o(n)$. In that case, random coding with Feinstein's suboptimal decoding, which suffices to achieve the capacity of conventional multiaccess channel capacity, can achieve the capacity of Gaussian MnAC. Proving the capacity achievability for faster scaling of the number of active users is much more challenging, mainly because the exponential number of possible error events prevent one from using the simple union bound. In this paper, we derive the capacity of MnAC for the case where the number of users may grow as quickly as linearly with the blocklength  by lower bounding the achievable error exponent based on maximum-likelihood decoding. The proposed MnAC model together with the capacity result and the compressed sensing based detection technique will provide insights for the optimal design in emerging applications such as machine-to-machine communication, where the number of devices in a cell may far exceed the blocklength. The results also complement a related study of many-broadcast models in~\cite{chen2014manybroadcast}.

Unless otherwise noted, we use the following notational conventions: $x$ denotes a scalar, $\bx$ denotes a vector and $\bxx$ denotes a matrix. The uppercase letters $X$, $\bX$ and $\bXX$ denote the corresponding random scalar, random vector and random matrix, respectively. Given a set $A$, $\bx_A$ and $\bxx_A$ denote the set of variables and the column vectors chosen from $\bx$ and $\bxx$ indexed by $A$, respectively. All logarithms are natural.


\section{System Model}
\label{sec:systemmodel}

The memoryless Gaussian many-access channel with on-off random access is defined as follows. Let $n$ denote the blocklength in the number of channel uses. The total number of users depends on $n$ and is explicitly denoted as $\ell_n$. In each block, each user accesses the channel independently with probability $\alpha_n$. The received symbols in a block is given as an $n$-vector:
\begin{equation}\label{eq:systemmodel}
\bY = \sum\limits_{k =1 }^{\ell_n} \bS_k (w_k) + \bZ,
\end{equation}
where $w_k$ is the message of user $k$, $\bS_k (w_k) \in \mathbb{R}^{n }$ is the corresponding codeword consisting of the $n$ transmitted symbols from user $k$, and $\bZ \in \mathbb{R}^{n}$ is the Gaussian noise vector with independent standard Gaussian entries.

\begin{definition}\label{def:MNCode}
Let $\mathcal{S}_k$ and $\mathcal{Y}$ denote the input alphabet of user $k$ and output alphabet, respectively. An $(M,n)$ code for the MnAC channel $(\mathcal{S}_1 \times \mathcal{S}_2 \times \cdots \times \mathcal{S}_{\ell_n}, p_{Y|S_1, \dots, S_{\ell_n}},\mathcal{Y})$ consists of the following mappings:
\begin{enumerate}
  \item Encoding functions $\mathcal{E}_k: \{ 0,1,\dots,M \} \rightarrow \mathcal{S}_k^n$ for $k = 1, \dots, \ell_n$, which maps the message $w$ to the codeword $\bs_k (w) =[s_{k1}(w), \cdots, s_{kn} (w) ]^T$. Every codeword $\bs_k (w)$ satisfies the following power constraint:
      \begin{equation} \label{eq:powerconst}
       \frac{1}{n} \sum\limits_{i=1}^n s_{ki}^2 (w) \leq P.
      \end{equation}
  If user $k$ is inactive in a given block, it is said to transmit the all-zero codeword $\bs_k(0) = \mathbf{0}$.
  \item Decoding function $\mathcal{D}: \mathcal{Y}^n \rightarrow \{0, 1,\dots,M \}^{\ell_n}$, which is a deterministic rule assigning a decision on the messages to each possible received vector.
\end{enumerate}
Suppose that the messages $w_1, \dots, w_{\ell_n}$ are sent by the $\ell_n$ users, respectively. Then the error probability is
\begin{align}
\nonumber \lambda_{\bm} &= P\left\{ \mathcal{D}(\bY) \neq (w_1, \dots, w_{\ell_n}) | \bs_1 = \mathcal{E}_1 (w_1), \right. \\
&  \left. \qquad \dots, \bs_{\ell_n} = \mathcal{E}_{\ell_n} (w_{\ell_n}) \right\}.
\end{align}
The average error probability for an $(M,n)$ code is:
\begin{equation}\label{eq:Perror}
P_e^{(n)} = P \left\{ \mathcal{D}(\bY) \neq (W_1, \dots, W_{\ell_n}) \right\},
\end{equation}
where $W_1, \cdots, W_{\ell_n} $ are independent, and for every $k \in \{1, \cdots, \ell_n\}$, $P\{W_k = 0 \} = 1- \alpha_n$ and $P\{W_k = w \} = \alpha_n / M$, for every $w = 1, \dots, M$.
\end{definition}

The preceding model reduces to the conventional $\ell$-user multiaccess channel in the special case where $\ell_n = \ell$ and $\alpha_n =1$.

\begin{definition}[Achievable message length sequence]\label{def:msglen}
We say a sequence of message lengths $\{ v(n) \}_{n=1}^{\infty}$ is asymptotically achievable for the MnAC if there exists a sequence of $(\lceil \exp(v(n)) \rceil,n)$ codes in the sense of Definition~\ref{def:MNCode} such that the error probability $P_e^{(n)}$ given by \eqref{eq:Perror} vanishes as $n \to \infty$.
\end{definition}

\begin{definition}[Symmetric capacity]\label{def:symmetriccapacity}
For the MnAC channel described by \eqref{eq:systemmodel}, $C(n)$ is said to be a symmetric capacity of the MnAC channel if for every $0 <\epsilon <1$, $(1 - \epsilon) C(n)$ is an asymptotically achievable message length whereas $(1+\epsilon)C(n)$ is not.
\end{definition}

For the special case of multiaccess channel, the $C(n)$ is essentially linear in $n$, so that $\lim_{n \to \infty} C(n) / n$ is equal to the conventional symmetric capacity of the multiaccess channel. In general, however, $C(n)$ need not grow linearly with the blocklength. Moreover, only the leading term of $C(n)$ matters in Definition~\ref{def:symmetriccapacity}. In particular, if $C(n)$ is a symmetric capacity, so is $C(n)+o(C(n))$.

For ease of analysis, we often use the following equivalent model for the Gaussian MnAC,
\begin{equation}\label{eq:systemmodel2}
\bY = \bSS \bX + \bZ,
\end{equation}
where $\bSS \in \mathbb{R}^{n \times M \ell_n}$ consists of the concatenated codebooks of the users, $\bZ \in \mathbb{R}^{n}$ is the Gaussian noise vector and $\bX \in \mathbb{R}^{M \ell_n}$ is a vector indicating the codewords transmitted by the users. Specifically, $\bX = [\bX_1^T, \bX_2^T, \cdots , \bX_{\ell_n}^T]^T$, where $\bX_k \in \mathbb{R}^M$ indicates the codeword transmitted by user $k$. For user $k$, $k=1,\cdots, \ell_n$, $\bX_k =  \mathbf{0}$ with probability $1 - \alpha_n$ and $\be_m$ with probability $\alpha_n / M$, where $\be_m$ is the binary vector with a single 1 at the $m$-th entry, $m = 1, \cdots, M$.

Note that $\bX$ must take its values in the following set:
\begin{align}
\nonumber   \mathcal{X}^{\ell_n}_M = & \left\{ \bx  =\left[ \bx_1^T, \cdots, \bx_{\ell_n}^T \right]^T : \text{for every } i \in \{1, \cdots, \ell_n \}, \right. \\
\label{eq:B} &  \quad \bx_i  = \mathbf{0} \text{ or } \be_j, \text{ for some } j=1, \cdots, M   \Big\}.
\end{align}

For notational convenience, further define $k_n  = \alpha_n \ell_n$ as the average number of active users. We focus on the regimes which satisfy the following two assumptions:

\noindent \textit{Assumption 1}: $k_n  = O(n)$ and the limit of $k_n$ exists.


\noindent \textit{Assumption 2}: If $k_n$ is unbounded, then $\ell_n e^{- \delta k_n } \to 0$ as $n \to \infty$ for any positive constant $\delta$.

\textit{Assumption 1} prohibits the uninteresting case where the average number of active users $k_n$ grows faster than linear in $n$.  For example, if $k_n=n (\log n)^2$, an average user will not be able to transmit a single bit reliably as $n$ increases to infinity. \textit{Assumption 2} disallows the growth rate of the total number of users $\ell_n$ to increase exponentially in $n$.

Time sharing with power allocation, which can achieve the capacity of the conventional multiaccess channel~\cite{cover06}, is inadequate for the MnAC in the regime of interest. For example, if $k=n$, each user would have only one channel use and cannot send even one bit reliably.

The following theorem is the main result of the paper.
\begin{theorem}[Capacity of MnAC with random user activity]\label{thm:capacityKMac}
For the MnAC channel described by \eqref{eq:systemmodel2}, the symmetric capacity $C(n)$ is characterized as follows,

\noindent (1) If $k_n  = \alpha_n \ell_n $ is unbounded, then
\begin{equation} \label{eq:symmcapacity}
C (n) = \left( \frac{n}{2 k_n } \log(1 + k_n  P) - \frac{H_2 (\alpha_n)}{\alpha_n} \right)^+ ,
\end{equation}
where $(x)^+  = \max(x,0)$.

\noindent (2) If $\ell_n$ is unbounded and $k_n$ is bounded, then $C(n) = o(n)$. Moreover, the message length $\frac{n}{2 s_n}  \log s_n $ is achievable for every positive unboundedly increasing $s_1, s_2, \dots$.

\noindent (3) If $\limsup_{n \to \infty} \ell_n = \ell_{0} < \infty$, then $C(n) = \frac{n}{2 \ell_{0} } \log(1 + \ell_{0}  P) $.
\end{theorem}

Case (3) can be easily proved by noticing that there is a non-vanishing probability that the number of active users is $\ell_{0}$, hence the capacity follows from the result for the conventional multiaccess channel with the maximum number of $\ell_0$ users. In the following, we focus on the case of unbounded $\ell_n$.

\begin{figure}
  \centering
  \includegraphics[width=5cm]{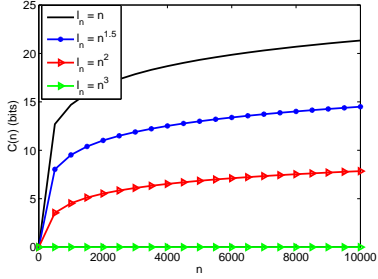}\\
  \caption{Plot of $C(n)$ given by \eqref{eq:symmcapacity}, where $P=2$, $k_n = n/4$.}\label{fig:Cn}
\end{figure}

Fig.~\ref{fig:Cn} illustrates the capacity $C(n)$ given
by~\eqref{eq:symmcapacity} in the special case where $P=2$ (i.e., SNR=3
dB), $k_n = n/4$, with different scalings of user number $\ell_n $.
The capacity (in message length) $C(n)$ does not
scale linearly in $n$.  Moreover, $C(n)$ depends on the scaling of
$k_n$ and $\ell_n$, whose effects cannot be captured by the
conventional multiaccess channels.
If $\ell_n$ grows too quickly (e.g., $\ell_n=n^3$), an average user
cannot transmit a single bit reliably. 

\section{Proof of the Converse of Theorem~\ref{thm:capacityKMac}}
\label{sec:converse}

We first prove the converse for unbounded $\ell_n$ and bounded $k_n$. Suppose $\limsup_{n \to \infty} C(n) / n = C$ for some $C>0$. There must exist some $k_0>0$ such that $\frac{1}{2 k_0} \log (1+ k_0 P) = C$. Then $C$ is the symmetric capacity of the conventional multiaccess channel with $k_0$ users. However, as $n \to \infty$, there is a non-vanishing probability that the number of active users is greater than $2 k_0$. Obviously, letting each user transmit a message length of $n C$ would yield a strictly positive error probability. By contradiction, $C(n) = o(n)$.

In the following, we provide the proof for the case of unbounded $k_n$ and $\ell_n$.

The transmitted signal $\bX$ in \eqref{eq:systemmodel2} is a binary vector, whose expected support size is $k_n $. Based on the input distribution described in Section~\ref{sec:systemmodel},
\begin{align}\label{eq:lbHB}
H(\bX) & = \ell_n H(X_1) = \ell_n (H_2(\alpha_n) + \alpha_n \log M) .
\end{align}

The detection error is equivalent to the erroneous estimation of $\bX$. Let $E = 1 \{ \hat{\bX} \neq \bX \} $, indicating if the estimation $\hat{\bX}$ is correct. Consider an $(M,n)$ code with a decoding function such that $P_e^{(n)} = P\{E = 1\}$ and it satisfies the power constraint \eqref{eq:powerconst}. For $0< \delta <1$, define 
\begin{equation}\label{eq:B1}
\mathcal{B}_{M}^{\ell_n} (\delta) = \left\{ \bx \in \mathcal{X}_M^{\ell_n}:  (1-\delta) k_n  \leq || \bx ||_0 \leq (1+ \delta) k_n  \right\},
\end{equation}
where $|| \cdot ||_q$ denotes the $q$-norm of a vector. The input entropy $H(\bX)$ can be calculated as
\begin{align}
H(\bX) & = H(\bX | \bY) + I(\bX; \bY) \\
\label{eq:HX} & = H\left(\bX, 1\left\{ \bX \in \mathcal{B}_M^{\ell_n} (\delta) \right\} \middle|  \bY \right) + I(\bX; \bY).
\end{align}
Applying the chain rule of entropy and using the fact that $E$ is determined by $\bX$ and $\bY$, we have
\begin{align}
\nonumber & H(\bX)  \\
\nonumber & =  H \left( 1 \left\{ \bX \in \mathcal{B}_M^{\ell_n} (\delta) \right\}  \middle| \bY \right) + H\left(  E \middle| \bY , 1 \left\{ \bX \in \mathcal{B}_M^{\ell_n} (\delta) \right\} \right )  \\
& \quad + H\left( \bX \middle| E, \bY, 1 \left\{ \bX \in \mathcal{B}_M^{\ell_n} (\delta) \right\} \right) + I(\bX; \bY) \\
\nonumber & \leq H_2 \left( P \left\{\bX \in \mathcal{B}_M^{\ell_n} (\delta) \right\} \right) + H_2 \left(P_e^{(n)}\right) + \\
\label{eq:Fanoub} & \qquad H \left( \bX \middle| E, \bY, 1 \left\{ \bX \in \mathcal{B}_M^{\ell_n} (\delta) \right\} \right) + I(\bX; \bY) ,
\end{align}
where $H_2(p) = - p \log p - (1-p) \log p$.

In order to derive a desired upper bound of the achievable message length, we use the following lemmas, whose proofs are omitted due to space limitations.
\begin{lemma}\label{lemma:hy}
$I(\bX ; \bY) \leq \frac{n}{2} \log \left(1 + k_n  P \right).$
\end{lemma}

\begin{lemma}\label{lemma:HBE}
For large enough $n$,
\begin{align}
\nonumber & H \left( \bX \middle| E, \bY, 1 \left\{ \bX \in \mathcal{B}_M^{\ell_n} (\delta) \right\} \right) \leq 4 P_e^{(n)} \left[ k_n  \log M \right. \\
&  \left. +  k_n  + \ell_n H_2 ( \alpha_n) \right] + \log M.
\end{align}
\end{lemma}

Note that
\begin{equation}
 H_2 \left( P_e^{(n)} \right)+ H_2 \left( P \left\{ \bX \in \mathcal{B}_M^{\ell_n} (\delta) \right\} \right)  \leq 2 \log 2
\end{equation}
and $k_n $ tends to infinity as $n$ increases. Combining~ \eqref{eq:lbHB}, \eqref{eq:HX}, \eqref{eq:Fanoub}, Lemma~\ref{lemma:hy} and Lemma~\ref{lemma:HBE}, we have for large enough $n$
\begin{align}
\nonumber &\left( 1 - 4 P_e^{(n)}  - \frac{1}{k_n } \right) \left( \log M + \frac{ H_2 ( \alpha_n)}{\alpha_n} \right)  \leq  \\
\label{eq:ublogM1} & \qquad \frac{n}{2 k_n } \log (1+ k_n  P)  + \delta + 4 P_e^{(n)} .
\end{align}
We further define several variables that are closely related to the upper bound of the message length. Let
\begin{equation}\label{eq:Cd}
C_1 (n) =  \frac{n}{2 k_n } \log(1 + k_n  P)  .
\end{equation}
Then $C(n)$ can be written as
\begin{equation}\label{eq:capacity}
C(n) = (1 - \theta_n)^+ C_1 (n),
\end{equation}
where
\begin{equation}\label{eq:gammaT}
\theta_n = \frac{H_2 (\alpha_n)}{\alpha_n C_1 (n)} = \frac{2 \ell_n H_2(\alpha_n)}{n \log (1+ \alpha_n \ell_n P)}.
\end{equation}

Since $P_e^{(n)} $ vanishes as $n$ increases and $\delta$ can be chosen arbitrarily small, given any $\epsilon >0$, there exists some $\delta$ and for large enough $n$ such that the following holds:
\begin{align}
 \log M & \leq (1 + \epsilon) C_1 (n) - \frac{H_2(\alpha_n)}{\alpha_n} \\
& = (1 + \epsilon - \theta_n) C_1 (n) \\
\label{eq:ublogM} & \leq \left( 1 + \frac{\epsilon}{1 - \theta_n} \right) C(n).
\end{align}

When $\theta_n <1$, $C(n)>0$ and \eqref{eq:ublogM} implies $\log M \leq (1+\epsilon) C(n)$ for any arbitrarily small $\epsilon$. When $\theta_n >1$, $C(n)=0$, meaning that an average user cannot send a single bit of information reliably through the Gaussian MnAC. The preceding argument shows that the symmetric capacity of MnAC with random access must not exceed $C(n)$.

A heuristic understanding of the expression of $C(n)$ in \eqref{eq:capacity} is as follows: If a genie-aided receiver reveals the set of active users to the receiver, the total number of bits that can be communicated through the MnAC with $k_n $ users would be approximately $(n/2) \log(1 + k_n  P)$, so that the symmetric capacity is $C_1(n)$. The capacity penalty on each of the $k_n $ active users is $H_2 (\alpha_n) / \alpha_n$ bits, because the total uncertainty in the activity of all $\ell_n$ users is $\ell_n H_2(\alpha_n) = k_n H_2(\alpha_n) / \alpha_n$.


\section{Proof of the Achievability of Theorem~\ref{thm:capacityKMac}}
\label{sec:achievability}

\begin{figure}
  \centering
  \includegraphics[width=7.5cm]{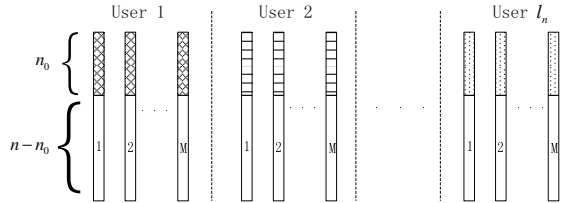}\\
  \caption{Codebook structure. Each user maintains $M$ codewords with each consisting of a message-bearing codeword prepended by a signature.}\label{fig:codebook}
\end{figure}

We first assume unbounded $k_n$ and establish an achievability result that is slightly stronger than that in Theorem~\ref{thm:capacityKMac}. The case of bounded $k_n$ is then straightforward.

We consider a two-stage approach: In the first stage the set of active users are identified based on their unique signatures and in the second stage the messages from the active users are decoded. Specifically, the following scheme is used:

\begin{itemize}

\item \textit{Codebook construction:} The codebooks of the $\ell_n$ users are generated independently. The codebook of user $k$ is generated as follows: Let
     \begin{equation}\label{eq:NT0}
     n_0 = \left \{
     \begin{array}{cc}
       \epsilon n , & \text{if  } \theta_n \to 0 \\
       \left( 1 + \frac{\epsilon}{\log ( 1+k_n  P) } \right) \theta_n n, & \text{otherwise.  }
     \end{array}
     \right.
     \end{equation}
     where $\epsilon$ is some positive number and $\theta_n$ is given by~\eqref{eq:gammaT}. The number of codewords $M$ is chosen such that
     \begin{equation}\label{eq:logCap1}
     \log M = \left\{
     \begin{array}{cc}
       (1 - \epsilon) C(n) \quad & \text{if  } \theta_n \to 0, \\
       C(n) - \epsilon n / k_n  \quad & \text{otherwise.  }
     \end{array}
     \right.
    \end{equation}
 The $w$-th codeword consists of two parts. The first $n_0$ symbols of each codeword contains the same signature $\bS_k^a \in \mathbb{R}^{n_0}$ that identifies the user. Each entry of the signature is generated according to i.i.d. $\mathcal{N} (0,P')$, $P' = P - \delta$. The remaining $n - n_0$ entries of the codeword $\bS_k(w)$ bear the message $w$. The $M(n - n_0)$ entries of all $M$ codewords are i.i.d. $\mathcal{N} (0,P')$. In other words, the $w$-th codeword of user $k$ is $\bS_k(w) = \left[ \left( \bS_k^a \right)^T \quad \left( \bS_k^b(w) \right)^T \right]^T$.
 The concatenated codebook matrix is illustrated in Fig.~\ref{fig:codebook}.

\item \textit{Transmission:} To send message $w_k$, user $k$ uses the first $n_0$ channel uses to transmit the signature $\bS_k^a$. The remaining $n- n_0$ channel uses are used to transmit the message-bearing codeword $\bS_k^b (w_k)$.
\item \textit{Channel: } A fraction of $\alpha_n$ users are active and transmit simultaneously. The received signal is $\bY$ given by \eqref{eq:systemmodel2}.
\item \textit{Two-stage detection and decoding:} Upon receiving $\bY$, the decoder performs the following:

(1) Activity identification: Let $\bY^a$ denote the first $n_0$ entries of $\bY$, corresponding to the superimposed signatures of the active user subject to noise, equivalently given by
      \begin{equation}\label{eq:identify}
      \bY^a = \bSS^a \bX^a + \bZ^a,
      \end{equation}
      where $\bX^a \in \mathbb{R}^{\ell_n}$ is a random vector with each entry following i.i.d. Bernoulli distribution with mean $\alpha_n$, and $\bZ^a \in \mathbb{R}^{n_0}$ is the Gaussian noise and $\bSS^a = [\bS^a_1 \cdots, \bS^a_{\ell_n}]$.

      The receiver searches, among all binary $\ell_n$-vectors whose support size does not exceed the average $k_n$ by too much, the activity vector that best explains the received signal:
      \begin{align}
        \label{eq:decode} \text{minimize} \quad & || \bY^a - \bss^a \bx ||_2^2 \\
        \nonumber \text{subject to }\quad &\bx \in \{0,1\}^{\ell_n} \\
        \nonumber \quad & ||\bx||_0 \leq (1 + 2 \delta_{n}) k_n  ,
        \end{align}
        where $\delta_n$ is some monotone decreasing sequence satisfying $\delta_{n}^2 k_n$ is unboundedly increasing and $\delta_n \log k_n  \to 0$.

(2) Message decoding: Let $\bY^b$ denote the last $n-n_0$ entries of $\bY$, corresponding to the superimposed message-bearing codewords. The receiver uses the maximum likelihood decoding to decode the messages based on the detected set of active users obtained from the first stage.
\end{itemize}


\begin{theorem}\label{thm:logCapStage1}
Using the decoding rule described by~\eqref{eq:decode} and $n_0$ channel uses, where $n_0$ is defined by \eqref{eq:NT0}, the user identities can be jointly determined with vanishing error probability as $n $ tends to infinity.
\end{theorem}

\begin{theorem}[Capacity of MnAC with known user activities]\label{thm:logCapStage2}
Consider $n$ channel uses of the MnAC. Suppose there are $k_n=O(n)$ active users and a genie reveals their identities to the receiver. Then the symmetric capacity is $C_1 (n) $ as defined by \eqref{eq:Cd}. In particular, there exists a sequence of codebooks with message lengths (in nats) $C_1 (n) (1 - \epsilon / \log (1+ k_n  P) ) $ such that the average error probability is arbitrarily small for sufficiently large $n$.
\end{theorem}

Let $n_0$ be defined by \eqref{eq:NT0}. Theorem~\ref{thm:logCapStage1} states that the identities of all active users can be detected with vanishing error probability using $n_0$ channel uses. Theorem~\ref{thm:logCapStage2} implies that, with the remaining $n-n_0$ channel uses, once the $k_n$ active users' identities have been revealed, the message length $ (1- \epsilon)\frac{ n-n_0 }{2 k_n} \log(1+k_n P)= (1- \epsilon)C(n)$ is achievable.  Hence the achievability of Theorem~\ref{thm:capacityKMac} is achieved.

Due to space limitations, we only provide a sketch of the proof of Theorems~\ref{thm:logCapStage1} and~\ref{thm:logCapStage2}. Essentially the same idea is used to bound the error probability in the two theorems.  Basically, the exponential number of error events are divided into a polynomial number of classes.  The error exponent for each class is characterized and shown to admit a universal positive lower bound.  Thus the total error probability vanishes.

We omit the treatment of Theorem~\ref{thm:logCapStage1} and focus on Theorem~\ref{thm:logCapStage2}. We can model the MnAC with known user activities as a special case of \eqref{eq:systemmodel2} with $\alpha_n = 1$, i.e., $k_n  = \ell_n$. The converse part of Theorem~\ref{thm:logCapStage2} follows naturally from Section~\ref{sec:converse} with $\alpha_n=1$. We focus on the achievability part. Upon receiving the length-$n$ vector $\bY$, we estimate $\bx = \left[\bx^T_1, \cdots, \bx^T_{k_n } \right]^T$ using the maximum likelihood decoding:
\begin{align}
\text{minimize} \quad &  || \bY - \bss \bx ||^2 \\
\text{subject to} \quad & \bx_k \in \left\{ e_1, \cdots, e_M \right\}, \quad \forall k=1, \dots, k_n.
\end{align}

Let $x^{\ast}$ denote the actual transmitted signal, whose support is denoted by $A^{\ast}$. Define $\mathcal{F}_k$ as the event that all the codewords of user $k$ satisfy the power constraint~\eqref{eq:powerconst}. Define $\mathcal{E}_{k}$ as the error event that $k$ users are received in error. The average error probability is upper bounded as
\begin{align}\label{eq:avgPe}
P_e^{(n)} \leq P\left\{ \mathcal{F}_1^c \cup \cdots  \mathcal{F}_{k_n }^c \right\} + \frac{1}{M^{k_n }} \sum\limits_{A^{\ast}}  \sum\limits_{ k=1}^{k_n } P\{\mathcal{E}_{k} | A^{\ast} \}.
\end{align}

Further denote $\gamma = k / k_n $ as the fraction of users subject to errors. Then we write $P\{\mathcal{E}_{k} | A^{\ast}  \}$ and $P\{\mathcal{E}_{\gamma}  | A^{\ast} \}$ interchangeably. In the following, we derive an achievable error exponent for the decoding error probability. The error exponent is closely related to the channel transition probability $p_{Y | \bS_A} ( y | \bs_A) $, i.e., the conditional distribution of $y$ given that the codewords $\bs_A = \{s_k: k \in A \}$ are transmitted.

\begin{lemma}
For an Gaussian MnAC with known user activities, there exists an $( \lceil \exp(v(n)) \rceil, n)$ code such that $P\{\mathcal{E}_{\gamma} | A^{\ast} \} \leq \exp \left[ - n f \left( \gamma, \rho \right) \right]$ for all $A^{\ast}$ and $\gamma$ with $1\leq \gamma k_n \leq k_n$,
where $\rho$ is any number with $0< \rho \leq 1$,
\begin{equation}
f \left( \gamma, \rho \right) = E_0(\gamma, \rho) - \frac{k_n }{n} H_2(\gamma) - \gamma \rho \frac{k_n }{n} v(n),
\end{equation}
and $E_0(\gamma, \rho)$ is defined as
\begin{equation} \label{eq:E0}
E_0(\gamma, \rho) = \frac{\rho}{2} \log \left( 1 +\frac{\gamma k_n  P' }{\rho+1 }  \right).
\end{equation}
\end{lemma}

It can be shown that $E_0 (\gamma, \rho)$ is an achievable error exponent for the error
probability caused by a particular $A$ being detected in favor of $A^{\ast}$, where $|A|=k_n$. The following lemma shows
some important properties of $ E_0(\gamma, \rho) $.
\begin{lemma}\label{lemma:E0}
The function $ E_0(\gamma, \rho) $ given by \eqref{eq:E0} satisfies
\\(P1) $ E_0(\gamma, \rho) $ is increasing in $\rho$.
\\(P2) $ E_0(\gamma, \rho) $ is concave in $\rho$.
\\(P3) $\frac{\partial E_0(\gamma, \rho)}{\partial \rho} |_{\rho =0}  = \frac{1}{2} \log \left( 1+ \gamma k_n P' \right)$.
\end{lemma}
\textit{Proof:} The proof follows similarly to \cite[Appendix 5B]{G68}.

The achievable error exponent for the overall detection error probability is thus essentially determined by
\begin{align}\label{eq:Er}
E_r = \min_{ \frac{1}{k_n } \leq \gamma \leq 1 }  \max_{0 \leq \rho \leq 1} f(\gamma, \rho) .
\end{align}
In order to prove that a message length $v(n)$ is achievable, it suffices to show that $n E_r \to \infty$ as $n \to \infty$. The following lemma is key to establishing Theorem~\ref{thm:logCapStage2}.
\begin{lemma}\label{lemma:errexp}
Let $M$ be such that the message length $v(n) = \log M$ is given by
\begin{equation}\label{eq:msglen}
v(n) = \frac{n}{2 k_n } (\log(k_n  P'  +1) - \epsilon) .
\end{equation}
Suppose $k_n = O(n)$, for sufficiently large $n$, $E_r$ given by \eqref{eq:Er} is greater than some fixed positive constant $d$.
\end{lemma}


Since Lemma \ref{lemma:errexp} holds for any $P' <P$ and any small enough $ \epsilon$, for large enough $n$, $P\{\mathcal{E}_{k} | A^{\ast} \} \leq e^{- d n}$. Then $\sum_{k=1}^{k_n } P\{\mathcal{E}_{k} | A^{\ast} \} \leq k_n  e^{- d n}$ vanishes. Moreover, with $v(n)$ specified by Lemma~\ref{lemma:errexp}, the probability of violating the power constraint vanishes as shown in~\cite{chen2013gaussian}. As a result, the overall error probability $P_e^{(n)}$ given by \eqref{eq:avgPe} vanishes and thus Theorem~\ref{thm:logCapStage2} is established.

With high probability the number of active users is no more than $(1+ \delta_n) k_n $, then Theorem~\ref{thm:logCapStage1} and Theorem~\ref{thm:logCapStage2} conclude that the message length $\frac{n - n_0}{2  (1+ \delta_n)  k_n } \left( \log \left( 1 +  (1+ \delta_n)  k_n  P \right) - \epsilon \right) $ is asymptotically achievable. Since $\delta_n$ vanishes and $\epsilon$ can be any arbitrarily small number, by replacing $n_0$ with \eqref{eq:NT0}, it is easy to find that $C(n) - \epsilon n /k_n $ and $(1- \epsilon) C(n) $ is achievable for the case of non-vanishing $\theta_n$ and vanishing $\theta_n$, respectively. Hence the achievability part of Theorem~\ref{thm:capacityKMac} is proved in the case of unbounded $k_n$.

The achievability result for bounded $k_n$ follows similarly because the number of active users is less than $s_n$ for any unboundedly increasing sequence $s_n$ as $n$ goes to infinity.

\bibliographystyle{IEEEtran}
\bibliography{IEEEabrv,randommanyaccess}

\end{document}